\documentclass[aps,prd,twocolumn,floatfix,nofootinbib,superscriptaddress]{revtex4-2}
\usepackage{graphicx}
\usepackage{amsmath,amssymb,mathtools,siunitx}
\usepackage{color}

\newcommand{\mnras}{Mon. Not. R. Astron. Soc}
\newcommand{\aap}{Astron. AstroPhys.}

\begin{document}
\title{Assessment of universal relations among second-order moments of relativistic stars via reformulated perturbation equations}
\author{Koutarou Kyutoku}
\affiliation{Department of Physics, Graduate School of Science, Chiba University, Chiba 263-8522, Japan}
\affiliation{Interdisciplinary Theoretical and Mathematical Sciences Program (iTHEMS), RIKEN, Wako, Saitama 351-0198, Japan}

\date{\today}

\begin{abstract}
 We assess the universal relations among second-order moments of relativistic stars, namely the moment of inertia, tidal deformability, and spin-induced quadrupole moment, via reformulated perturbation equations. After constructing the spherical background configuration by solving two ordinary differential equations as usual, these three moments are obtained by solving four additional ordinary differential equations. They are solved numerically from the stellar center to the surface, and we do not need to derive homogeneous solutions for obtaining the quadrupole moment. This small number of ordinary differential equations to be solved enables us to identify the primary variable for each second-order moment. Investigating the profile of these variables in the star, we speculate that their nonmonotonic behavior, enhanced typically for soft equations of state and/or high compactnesses, introduces the variety to the relations among these second-order moments unless the black-hole limit is approached. Because realistic relativistic stars are widely believed to be characterized by stiff equations of state, they enjoy the universal relation to a great extent.
\end{abstract}

\maketitle

\section{Introduction} \label{sec:intro}

The orbital dynamics of binaries are affected by various second-order moments of the stellar density distribution on top of the point-particle interaction. These moments as functions of mass are determined by the equation of state for stellar matter. Thus, their measurement in neutron-star binaries enables us to understand the properties of supranuclear-density matter \cite{1992ApJ...398..569L,Harada:2001nz}. The moment of inertia $I$ relates the rotational period to the spin angular momentum and may be constrained by future observations of binary pulsars \cite{Lattimer:2004nj}. Tidal deformability $\Lambda$ affects the gravitational waveform in the inspiral phase at the effective\footnote{Here, ``effective'' means that it is really an effect in Newtonian gravity with the corresponding dependence on the orbital separation.} fifth post-Newtonian order \cite{Flanagan:2007ix} and has already been measured in GW170817 \cite{LIGOScientific:2018cki}. The spin-induced quadrupole moment $Q$ also affects the gravitational waveform at the effective second post-Newtonian order \cite{Poisson:1997ha} and could be important for rotating neutron stars as well as Kerr black holes. The quadrupole moment may also play a role in shaping X-ray emission such as pulse profiles from rapidly rotating neutron stars \cite{Psaltis:2013zja}.

It has been pointed out that the relations among $I$, $\Lambda$, and $Q$ of neutron (and quark) stars depend only very weakly on the equation of state under appropriate normalization \cite{Yagi:2013bca,Yagi:2013awa}. The origin of these so-called universal relations is not clearly understood yet. One possible clue to this question is that the universality tends to be violated for soft equations of state that are relevant to normal stars and white dwarfs \cite{Yagi:2013awa}.

Irrespective of its physical understanding, the universality for neutron stars is particularly useful for reducing the number of independent parameters in astrophysical inference. For example, along with the expectation that neutron stars in merging binaries may not be rapidly rotating, $Q$ is frequently expressed as a function of $\Lambda$ in gravitational-wave data analysis (see, e.g., Ref.~\cite{LIGOScientific:2018cki}). Interestingly though, even when the analysis pipeline directly samples the equation of state, $Q$ is sometimes computed indirectly from $\Lambda$ \cite{LIGOScientific:2018cki}.

The computation of the quadrupole moment is sometimes considered to be more complicated than it actually is. After solving the background and lower-order perturbation, the standard procedure integrates four ordinary differential equations. Two of them derive a set of particular solutions for metric perturbations usually denoted by $h_2$ and $v_2$, and the other two a set of homogeneous solutions \cite{Hartle:1967he}. These two sets of solutions are combined to ensure that $h_2$ and $v_2$ share an identical integration constant, which is needed to express the quadrupole moment. As a matter of fact, this (or equivalent) procedure is necessary if our purpose is to describe the spacetime and stellar structure in full detail at the relevant order of perturbation.

In this paper, we show that it is sufficient to solve two ordinary differential equations without explicitly deriving homogeneous solutions, if our purpose is only to obtain the quadrupole moment. Our formulation derives $I$, $\Lambda$, and $Q$ by solving only six ordinary differential equations. An additional benefit of this simplification is that the primary variable determining the quadrupole moment, at least its plausible candidate, may be identified.

This paper is organized as follows. We describe our procedure for obtaining the quadrupole moment in Sec.~\ref{sec:formulation}. Results for polytropic equations of state are presented in Sec.~\ref{sec:result} to demonstrate the validity of our procedure, followed by the assessment of universal relations among second-order moments focusing on corresponding primary variables. Section \ref{sec:summary} is devoted to a summary. Throughout this paper, the gravitational constant $G$ and the speed of light $c$ are set to unity.

\section{Differential equation for relativistic slowly rotating stars} \label{sec:formulation}

We first summarize equations for solving slowly rotating stars \cite{Hartle:1967he,Hartle:1968si} focusing on obtaining second-order moments. The rotational angular velocity is denoted by $\Omega$, and the equations on the order of $\Omega^0$, $\Omega^1$, and $\Omega^2$ are presented. The novel aspect of our reformulation lies in Eqs.~\eqref{eq:ode_f} and \eqref{eq:ode_q}, with which the spin-induced quadrupole moment is obtained by solving two ordinary differential equations from the center to the surface once and for all. Moreover, only one of them, Eq.~\eqref{eq:ode_q}, directly governs the quadrupole moment, with the other, Eq.~\eqref{eq:ode_f}, serving as an auxiliary condition. This feature enables us to specify the primary variable determining the quadrupole moment and to investigate its profile in the star.

\subsection{Zeroth order: TOV equation}

To fix the notation, we present equations governing nonrotating relativistic stars. We write the spacetime metric as\footnote{Whether we multiply $\nu$ and $\lambda$ by $2$ is largely a matter of taste. A possible advantage of the current choice is that $\nu$ asymptotes to the Newtonian potential, making the interpretation of Bernoulli's integral a bit simpler (see below).}
\begin{equation}
 \mathrm{d}s^2 = - e^{2\nu} \mathrm{d}t^2 + e^{2\lambda} \mathrm{d}r^2 + r^2 \left(\mathrm{d}\theta^2 + \sin^2 \theta \mathrm{d}\varphi^2 \right)
\end{equation}
and the energy-momentum tensor for the perfect fluid as
\begin{equation}
 T_\alpha{}^\beta = (\varepsilon + P) u_\alpha u^\beta + P \delta_\alpha{}^\beta ,
\end{equation}
where $u^\alpha$, $\varepsilon$, and $P$ are the fluid four velocity, energy density, and pressure, respectively.\footnote{Although it is also common to express the energy density by $\rho$, we refrain from this convention to avoid possible confusion with the rest-mass density.} Throughout this paper, the equation of state is assumed to be barotropic so that $\varepsilon$ is uniquely determined by $P$. The logarithm of the specific enthalpy (i.e., enthalpy per mass) given by
\begin{equation}
 H(P) = \int_0^P \frac{\mathrm{d}P'}{\varepsilon (P') + P'} \label{eq:def_h}
\end{equation}
is another useful thermodynamic quantity \cite{1992ApJ...398..569L}.

The stellar structure is governed by Tolman-Oppenheimer-Volkoff equation (see, e.g., Sec.~6.2 of Ref.~\cite{Wald:1984rg}),
\begin{equation}
 \frac{\mathrm{d}P}{\mathrm{d}r} = - \frac{(\varepsilon +P) (m+4\pi P r^3)}{r(r-2m)} , \label{eq:ode_p}
\end{equation}
along with the definition of mass,
\begin{equation}
 \frac{\mathrm{d}m}{\mathrm{d}r} = 4\pi \varepsilon r^2 , \label{eq:ode_m}
\end{equation}
which also gives
\begin{equation}
 e^{2\lambda} = \left( 1 - \frac{2m}{r} \right)^{-1} .
\end{equation}
Once a central value of the pressure $P_c = P(r=0)$ is specified, they are solved with $m(r \to 0) = (4\pi/3) \varepsilon_c r^3$, where $\varepsilon_c = \varepsilon (P_c)$, until the stellar surface defined by $P=0$ is encountered. This procedure gives us the stellar radius $R$ and the gravitational mass of the star $M = m(r=R)$ for a given value of $P_c$. We also introduce the compactness $C \coloneqq M/R$ for convenience.

As pointed out in Ref.~\cite{1992ApJ...398..569L}, it is often advantageous to convert the independent variable from $r$ to thermodynamic variables such as $H$, mainly because the location of the stellar surface where $H=0$ is predetermined. The conversion is performed by multiplying
\begin{equation}
 \frac{\mathrm{d}r}{\mathrm{d}H} = - \frac{r(r-2m)}{m+4\pi P r^3} , \label{eq:ode_r}
\end{equation}
which is obtained by combining Eqs.~\eqref{eq:def_h} and \eqref{eq:ode_p}. When this conversion is done, we directly solve this equation instead of Eq.~\eqref{eq:ode_p}. The same conversion can also be applied to perturbation equations explained in the following subsections.

The remaining metric component, $\nu$, is obtained algebraically via the relativistic Bernoulli's theorem (see, e.g., Sec.~3.4.2 of Ref.~\cite{Gourgoulhon:2010ju}) as
\begin{equation}
 \nu = \frac{1}{2} \ln (1 - 2C) - H .
\end{equation}
Although some literature seems to solve an additional differential equation
\begin{equation}
 \frac{\mathrm{d}\nu}{\mathrm{d}r} = - \frac{\mathrm{d}H}{\mathrm{d}r} = \frac{m+4\pi P r^3}{r(r-2m)} , \label{eq:ode_nu}
\end{equation}
this is unnecessary. Meanwhile, Eq.~\eqref{eq:ode_nu} is sometimes useful for expressing the differential equations for perturbation in a concise manner.

\subsection{First order: Moment of inertia}

The moment of inertia is obtained by solving the first-order perturbation in terms of $\Omega$ \cite{Hartle:1967he,Hartle:1968si}. The original procedure solves a second-order differential equation for $\overline{\omega} (r) \coloneqq \Omega - \omega (r)$, where $\omega$ is tightly related to the $t$-$\varphi$ component of the metric perturbation. Because the solution outside the star derives $\omega (r=R) = \Omega - 2J/R^3$ with $J \propto \Omega$ being the spin angular momentum of the star, the surface value of $\overline{\omega}$ gives us the moment of inertia $I \coloneqq J/\Omega$.

As demonstrated by Ref.~\cite{1987A&A...172...95Z}, this procedure is simplified by using the logarithmic derivative
\begin{equation}
 w(r) \coloneqq \frac{\mathrm{d} \ln \overline{\omega}}{\mathrm{d} \ln r} .
\end{equation}
First, the rotational angular velocity $\Omega$ scales out from the problem. The moment of inertia is given in terms of the surface value $w_s = w(r=R)$, and specifically, the normalized counterpart $\overline{I} \coloneqq I/M^3$ is expressed as
\begin{equation}
 \overline{I} = \frac{w_s}{C^3 (2w_s + 6)} . \label{eq:alg_i}
\end{equation}
Second, this function obeys a single first-order differential equation
\begin{equation}
 \frac{\mathrm{d}w}{\mathrm{d}r} = - \frac{w(w+3)}{r} + \frac{4\pi (w+4) (\varepsilon +P) r^2}{r-2m} . \label{eq:ode_w}
\end{equation}
Numerical integration is performed with the regularity condition $w(r \to 0) = (16\pi/5) (\varepsilon_c + P_c) r^2$.

Equation \eqref{eq:ode_w} suggests that $w$ serves as the primary variable determining the moment of inertia. Of course, this is not at all the unique choice of the primary variable. Rather, it is well known that the moment of inertia is expressed directly as the radial integral of $(8\pi/3) r^4 e^{\lambda - \nu} (\varepsilon +P) (\overline{\omega}/\Omega)$ \cite{Hartle:1967he}. Here, considering that the same approach appears inapplicable to other second-order moments, we focus on $w$ in this study.

\subsection{Second order: Quadrupole moment}

The quadrupole moment is obtained by solving the second-order perturbation in $\Omega$, which introduces the stellar deformation \cite{Hartle:1967he,Hartle:1968si}. Equations to determine the spin-induced quadrupole moment $Q$ are written schematically as
\begin{equation}
 \frac{\mathrm{d}}{\mathrm{d}r}
  \begin{pmatrix}
  h_2 \\ v_2
  \end{pmatrix}
  =
  \begin{pmatrix}
  L_{hh} & L_{hv} \\ L_{vh} & 0
  \end{pmatrix}
  \begin{pmatrix}
  h_2 \\ v_2
  \end{pmatrix}
  +
  \begin{pmatrix}
  W_h \\ W_v
  \end{pmatrix}
  .
\end{equation}
The matrix components $L_{hh}$, $L_{hv}$, and $L_{vh}$ are functions of background quantities, while the vector components $W_h$ and $W_v$ are functions of background quantities multiplied by $\overline{\omega}^2$ or $(\mathrm{d} \overline{\omega}/\mathrm{d}r)^2$. Their exact expressions can be found in, e.g., Eq.~(21) of Ref.~\cite{Hartle:1968si}. The regularity condition at the center requires $h_2 (r \to 0) = A r^2$ and $v_2 (r \to 0) = B r^4$ with a single constraint
\begin{equation}
 \frac{2\pi}{3} (\varepsilon_c + 3P_c) A + B = \frac{2\pi}{3} (\varepsilon_c + P_c) \overline{\omega}_c^2 e^{-2\nu_c} , \label{eq:reg}
\end{equation}
where $\overline{\omega}_c = \overline{\omega} (r=0)$ and $\nu_c = \nu (r=0)$. The exterior solution derives
\begin{align}
 v_2 (r=R) & = \frac{2K}{\sqrt{\zeta_s^2 -1}} Q_2^1 (\zeta_s) - \frac{J^2}{R^4} , \\
 h_2 (r=R) & = K Q_2^2 (\zeta_s) + \frac{J^2}{MR^3} + \frac{J^2}{R^4} ,
\end{align}
where $\zeta_s \coloneqq R/M - 1 = 1/C - 1\,(>1)$, $Q_l^m$ denotes the associated Legendre polynomials of the second kind [see Eqs.~(137) and (141) of Ref.~\cite{Hartle:1967he} for explicit forms], and $K$ is an integration constant. The quadrupole moment in the sign convention of Ref.~\cite{Hartle:1967he} is written in terms of this integration constant by
\begin{equation}
 Q = \frac{J^2}{M} + \frac{8KM^3}{5} .
\end{equation}

When we solve these equations from the stellar center, it is not straightforward to ensure that $h_2$ and $v_2$ end up in the same values of $K$ at the surface. This could have been accomplished by eliminating one of these two variables, and actually this can be done outside the star where $L_{hv} (r>R) = 2/M = \mathrm{const}.$ However, this is unlikely to be feasible inside the star. The usual procedure is to (i) derive one set of particular solutions starting from an arbitrary set of values for $A$ and $B$ satisfying Eq.~\eqref{eq:reg}, (ii) derive one set of homogeneous solutions setting $\overline{\omega} = 0$, and (iii) combine them to find the set of physical solutions with identical values of $K$. Leaving aside the computational burden, this does not give an insight into which differential equation governs the value of $K$ and hence $Q$.

If our purpose is only to obtain the quadrupole moment, we may introduce a new variable
\begin{equation}
 q(r) = v_2 (r) + f(r) h_2 (r) ,
\end{equation}
which obeys
\begin{equation}
 \frac{\mathrm{d}q}{\mathrm{d}r} = \frac{\mathrm{d}v_2}{\mathrm{d}r} + f \frac{\mathrm{d}h_2}{\mathrm{d}r} + \frac{\mathrm{d}f}{\mathrm{d}r} h_2 .
\end{equation}
By imposing that the right-hand side depends on $h_2$ and $v_2$ only via $q$, the auxiliary function $f$ is fixed to a solution of
\begin{align}
 \frac{\mathrm{d}f}{\mathrm{d}r} & = - \frac{2}{m+4\pi Pr^3} f^2 \notag \\
 & + \left\{2 \frac{\mathrm{d}\nu}{\mathrm{d}r} - \frac{r^2}{m+4\pi P r^3} \left[4\pi (\varepsilon +P) - \frac{2m}{r^3} \right] \right\} f + 2 \frac{\mathrm{d}\nu}{\mathrm{d}r} \label{eq:ode_f} .
\end{align}
We would also like to impose the condition at the center
\begin{equation}
 f (r \to 0) = \frac{2\pi}{3} (\varepsilon_c + 3P_c) r^2 ,
\end{equation}
so that the regularity condition Eq.~\eqref{eq:reg} directly applies to $q(r)$, specifically,
\begin{equation}
 q (r \to 0) = \frac{2\pi}{3} (\varepsilon_c + P_c) \overline{\omega}_c^2 e^{-2\nu_c} r^4 .
\end{equation}
These requirements uniquely determine $f(r)$ and hence $q(r)$.

We may further simplify the problem by changing the variable to $\tilde{q} \coloneqq q / (\overline{\omega} r)^2$, with which $\Omega$ again scales out from the problem. This dimensionless variable is determined by solving
\begin{align}
 \frac{\mathrm{d}\tilde{q}}{\mathrm{d}r} & = - \left[ \frac{2f}{m+4\pi Pr^3} + \frac{2(w+1)}{r} \right] \tilde{q} \notag \\
 & + \frac{j^2 w^2}{6} \left\{ \frac{\mathrm{d}\nu}{\mathrm{d}r} + \frac{1}{r} + \left[ \frac{\mathrm{d}\nu}{\mathrm{d}r} - \frac{1}{2(m+4\pi Pr^3)} \right] f \right\} \notag \\
 & + \frac{8\pi (\varepsilon +P) r^3 j^2}{3(r-2m)} \left\{\frac{\mathrm{d}\nu}{\mathrm{d}r} + \frac{1}{r} + \left[\frac{\mathrm{d}\nu}{\mathrm{d}r} + \frac{1}{2(m+4\pi Pr^3)} \right] f \right\} , \label{eq:ode_q}
\end{align}
where $j \coloneqq e^{-(\nu + \lambda)}$, starting from the regularity condition
\begin{equation}
 \tilde{q} (r \to 0) = \frac{2\pi}{3} (\varepsilon_c + P_c) e^{-2\nu_c} r^2 .
\end{equation}
The integration constant $K$ is readily determined by the surface values $\tilde{q}_s = \tilde{q} (r=R)$ and $f_s = f(r=R)$. In particular, the normalized quadrupole moment $\overline{Q} \coloneqq Q/(\chi^2 M^3)$, where $\chi \coloneqq J/M^2$ is the spin parameter, is expressed as
\begin{align}
 \overline{Q} & = 1 + \frac{8}{5} \left[ \frac{2Q_2^1 (\zeta_s)}{\sqrt{\zeta_s^2 - 1}} + f_s Q_2^2 (\zeta_s) \right]^{-1} \notag \\
 & \times \left\{\frac{\left(1 - 2C^3 \overline{I}\right)^2}{C^2 \overline{I}^2} \tilde{q}_s + C^4 \left[1 - \left(1 + \frac{1}{C} \right) f_s \right] \right\} . \label{eq:alg_q}
\end{align}
While this expression depends not only on $\tilde{q}_s$ but also on $f_s$ due to the obvious dependence of $q$ on $f$, we regard $\tilde{q}$ as the primary variable that determines the value of $K$ and hence $Q$. The reason for this includes the (interdependent) facts that (i) $q$ is directly related to the metric perturbations $v_2$ and $h_2$, (ii) $\mathrm{d}\tilde{q}/\mathrm{d}r$ includes nonlinear source terms $w^2$, and (iii) $\mathrm{d}\tilde{q}/\mathrm{d}r$ includes $\mathrm{d}f/\mathrm{d}r$. In fact, Eq.~\eqref{eq:alg_q} also depends on $w_s$ via $\overline{I}$, rendering unambiguous identification of the key variable for $\overline{Q}$ difficult.

\section{Numerical result} \label{sec:result}

\subsection{Relation among the second-order moments} \label{sec:result_ql}

\begin{figure*}[tbp]
 \includegraphics[width=.95\linewidth]{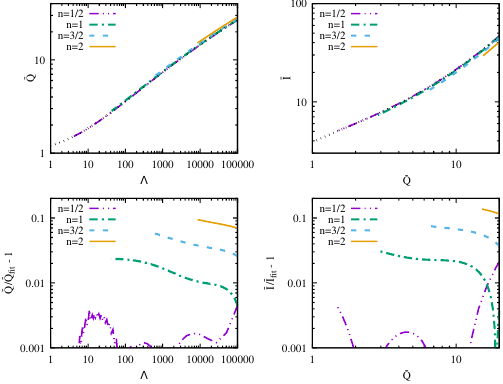}
 \caption{Relation between the normalized quadrupole moment $\overline{Q}$ and the dimensionless tidal deformability $\Lambda$ (top left) and the normalized moment of inertia $\overline{I}$ and $\overline{Q}$ (top right) for various polytropes. All the sequences are terminated at the maximum-mass configuration at the bottom left. This is essentially a reproduction of Fig.~1 (right) and Fig.~10 from Ref.~\cite{Yagi:2013awa}, which also derives the fitting formula shown in this plot (black dotted curve). The bottom panels show the relative deviation between the actual results and the fitting formula. Because our polytropic equations of state are defined in terms of the rest-mass density, the results are slightly different from those presented in Ref.~\cite{Yagi:2013awa}, in which the energy density is used.} \label{fig:lq}
\end{figure*}

To validate our procedure, we compute second-order moments for polytropic equations of state in terms of the rest-mass density (not the energy density) characterized by polytropic indices $n=1/2$, $1$, $3/2$, and $2$. In terms of the adiabatic index $\Gamma = 1 + 1/n$, they correspond to $\Gamma = 3$, $2$, $5/3$, and $3/2$, respectively. Typical nuclear-theory-based equations of state for neutron-star matter may be approximated by $n=1/2$--$1$ \cite{Yagi:2013awa}, and those for white dwarfs with nonrelativistic electron degeneracy are characterized by $n=3/2$. We adopt $n=2$ as a substitute for $n=3$ that characterizes massive white dwarfs with relativistic electron degeneracy to fit all the results into a single plot (see Fig.~\ref{fig:lq}). Thus, our choice of polytropic indices reasonably covers realistic compact objects.

For completeness, we reproduce the equations for determining quadrupolar tidal deformability \cite{Hinderer:2007mb}. Following Ref.~\cite{Postnikov:2010yn}, we again solve a single ordinary differential equation
\begin{align}
 \frac{\mathrm{d}y}{\mathrm{d}r} & = - \frac{y^2}{r} + \frac{e^{2\lambda}}{r} \left[ - 1 + 4\pi r^2 (\varepsilon -P) \right] y \notag \\
 & + r \left[ 4 \left( \frac{\mathrm{d}\nu}{\mathrm{d}r} \right)^2 + \frac{6 e^{2\lambda}}{r^2} - 4\pi e^{2\lambda} \left( 5\varepsilon + 9P + \frac{\varepsilon +P}{\mathrm{d}P/\mathrm{d}\varepsilon} \right) \right] \label{eq:ode_y}
\end{align}
for the logarithmic derivative $y$ of the relevant metric perturbation with the regularity condition $y(r=0) = 2$. The dimensionless tidal deformability $\Lambda$ is expressed in terms of its surface value $y_s = y(r=R)$ as
\begin{align}
 \Lambda & = \frac{16}{15} D^2 [2-y_s+2C(y_s-1)] \notag \\
 & \times \{2C [6-3y_s + 3C(5y_s-8)] \notag \\
 & + 4C^3 [13-11y_s + C(3y_s-2) + 2C^2 (1+y_s)] \notag \\
 & + 3D^2 [2-y_s+2C(y_s-1)] \ln D\}^{-1} , \label{eq:alg_l}
\end{align}
where $D \coloneqq 1 - 2C$. These equations suggest that $y$ serves as the primary variable determining $\Lambda$. Because Eq.~\eqref{eq:ode_y} involves $\mathrm{d}P/\mathrm{d}\varepsilon$, the square of the sound velocity, special care must be taken in the presence of density discontinuity such as the first-order phase transition \cite{Postnikov:2010yn}.

Figure \ref{fig:lq} shows $\overline{Q}$ as a function of $\Lambda$ and $\overline{I}$ as a function of $\overline{Q}$ for various polytropes obtained with our procedure. Equivalent results have already been presented in many places. Thus, this figure validates our simplified procedure for computing the spin-induced quadrupole moment. We also show the fitting formula for the claimed universal relation \cite{Yagi:2013awa}. As described in the previous study, stiff equations of state with a small value of $n$ (large $\Gamma$) nicely agree with the fitting formula, and the deviation becomes large as the equation of state becomes soft with a large value of $n$ (small $\Gamma$). We recall that, for typical neutron stars with $\Lambda \approx 100$--$1000$ and more massive ones, the universality holds within about percent accuracy.

\subsection{Comparison of $C=0.05$ configurations} \label{sec:result_C005}

\begin{figure*}[tbp]
 \begin{tabular}{cc}
  \includegraphics[width=.48\linewidth]{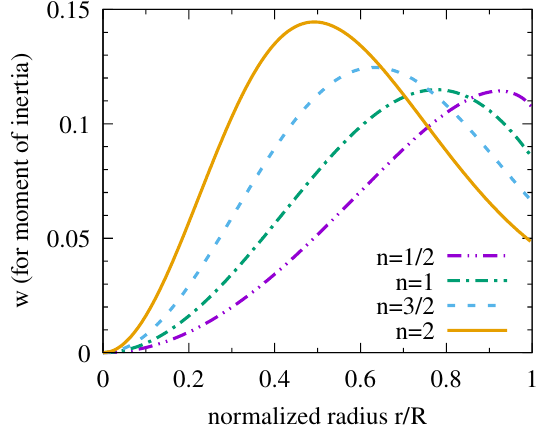} &
      \includegraphics[width=.48\linewidth]{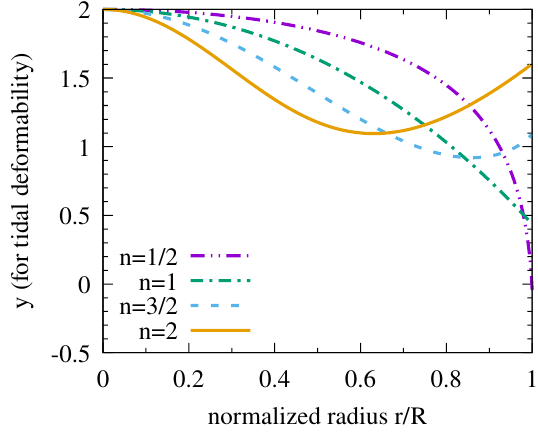} \\
  \includegraphics[width=.48\linewidth]{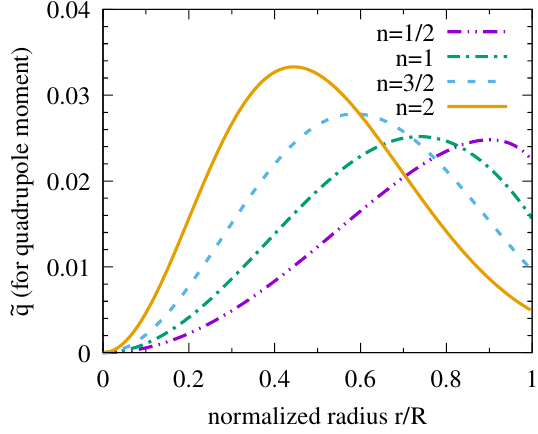} &
      \includegraphics[width=.48\linewidth]{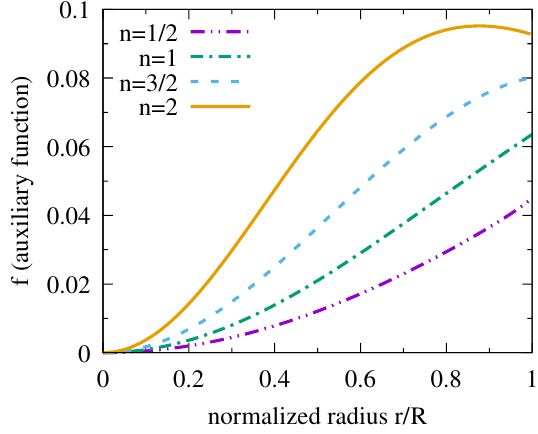}
 \end{tabular}
 \caption{Radial profile of $w$ (top left), $y$ (top right), $\tilde{q}$ (bottom left), and $f$ (bottom right) for the $C=0.05$ configurations. All the profiles are expressed as functions of the normalized radius, $r/R$.} \label{fig:C005}
\end{figure*}

To investigate the universality and deviation of second-order moments associated with the stiffness of the equation of state, we compare the radial profiles of $w$, $y$, $\tilde{q}$, and $f$ for the configurations with $C=0.05$ in Fig.~\ref{fig:C005}. This relatively small value is chosen because the compactness for the maximum-mass configuration is only $C = 0.074$ for $n=2$. The configurations with $C=0.05$ span from $\Lambda \approx \num{9e4}$ for $n=2$ to $\approx \num{7e5}$ for $n=1/2$.

As the equation of state becomes soft with a large value of $n$, the peaks of all these functions shift inward, presumably related to the applicability range of the universal relations. On the one hand, the rapid increase (or decrease in the case of $y$) in the central region for large $n$ is naturally expected from the centrally condensed density profile for a soft equation of state. On the other hand, it is not necessarily expected that these functions turn around toward the outer region. Indeed, $y$ and $f$ are monotonic for stiff polytropes with $n \le 1$. Taking into account that the universality tends to be violated for soft equations of state \cite{Yagi:2013awa}, we speculate that the decreasing part of these functions in the outer region introduces the variety to the relations among $\overline{I}$, $\Lambda$, and $\overline{Q}$, for better or worse.

Stated differently, universality appears to be approached as the characteristic length scales of these primary variables become long compared to the stellar radius. This description might align with the idea of universality in other branches of physics. That is, universality is achieved as the relevant length scales become infinite, e.g., the correlation length in critical phenomena or the scattering length in quantum few-body systems. However, as described in Sec.~\ref{sec:result_Max}, this does not explain the universality of high-compactness configurations.

While both $f$ and $\tilde{q}$ depend positively on the value of $n$ in the central region, the surface values $f_s$ and $\tilde{q}_s$ depend positively and negatively on $n$, respectively. The positive dependence of $f_s$ on $n$ is understood from the Newtonian limit; Eq.~\eqref{eq:ode_f} for $f$ reduces to (Clairaut-)Radau's equation of degree $l=2$ for $\eta \coloneqq 2rf/m - 1$, whose surface value is larger for larger $n$ (see, e.g., Ref.~\cite{1955MNRAS.115..101B} where our $l$ is denoted by $j$). This explains the positive dependence of $f_s$ and also its order of magnitude, $\mathcal{O}(C)$. In addition, this Newtonian limit may be contrasted with that of Eq.~\eqref{eq:ode_y}. The latter involves the square of the sound velocity, $\mathrm{d}P/\mathrm{d}\varepsilon$, and thus is not directly related to Radau's equation \cite{Hinderer:2007mb}.

The overall profile of $\tilde{q}$ is more similar to $w$ than to $f$ despite their explicit linear dependence. This is because, again in the Newtonian limit, $\mathrm{d}\tilde{q}/\mathrm{d}r = (8\pi \varepsilon r/3) [1 + rf/(2m)] - (2\tilde{q}/r)(1 + rf/m)$ shares a similar structure with $\mathrm{d}w/\mathrm{d}r = 16\pi \varepsilon r - 3w/r$. Once the density $\varepsilon$ drops sufficiently in the outer region, these functions begin to decrease by their own nature. Thus, the surface values $w_s$ and $\tilde{q}_s$ are decreased for softer equations of state due to the more rapid decrease in the density. These Newtonian limits also imply that $w_s$ and $\tilde{q}_s$ are on the order of $\mathcal{O} (C)$.

Although the surface values $f_s$ are generally larger than $\tilde{q}_s$, this does not mean that $f_s$ makes a more significant contribution to $\overline{Q}$. Rather, the first term in the second line of Eq.~\eqref{eq:alg_q} involving $\tilde{q}_s$ is larger by a few orders of magnitude than the second term involving $f_s$ due to the dependence on $C$ and $\overline{I}$. This may justify our expectation that $\tilde{q}_s$ is the primary variable determining the quadrupole moment. As for the denominator, the square bracket in the first line of Eq.~\eqref{eq:alg_q}, the second term $f_s Q_2^2 (\zeta_s)$ is larger by only a factor of a few than the first term.

\subsection{Comparison of maximum-mass configurations} \label{sec:result_Max}

\begin{figure*}[tbp]
 \begin{tabular}{cc}
  \includegraphics[width=.48\linewidth]{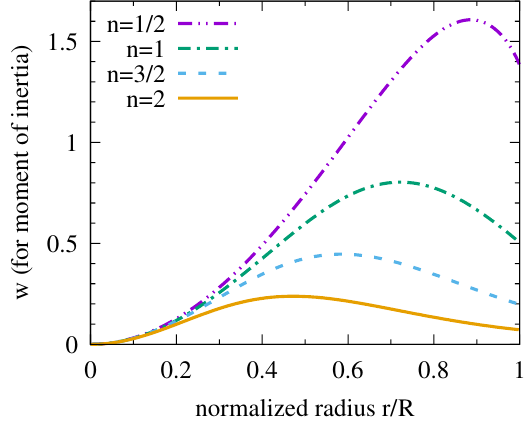} &
      \includegraphics[width=.48\linewidth]{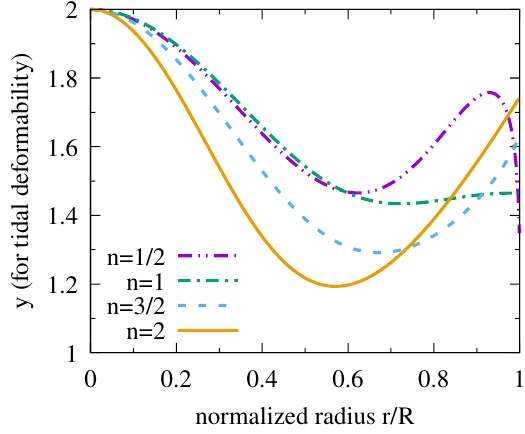} \\
  \includegraphics[width=.48\linewidth]{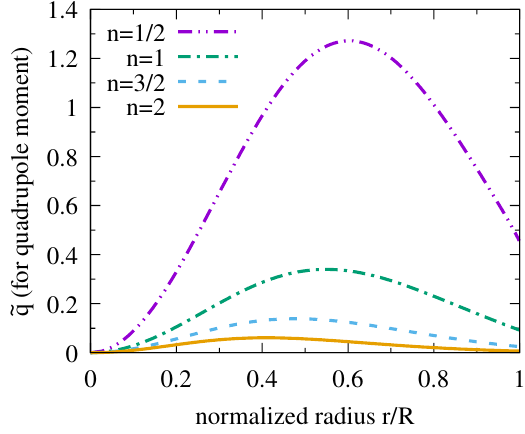} &
      \includegraphics[width=.48\linewidth]{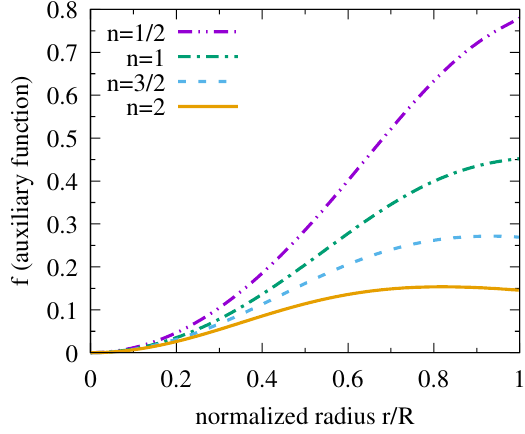}
 \end{tabular}
 \caption{Same as Fig.~\ref{fig:C005} but for the maximum-mass configurations. The compactness values are $C=0.32$, $0.21$, $0.135$, and $0.074$ for $n=1/2$, $1$, $3/2$, and $2$, respectively.} \label{fig:Max}
\end{figure*}

We also compare the radial profile of functions $w$, $y$, $\tilde{q}$, and $f$ for maximum-mass configurations in Fig.~\ref{fig:Max} to investigate the influence of strong gravity. The compactnesses are $C=0.32$, $0.21$, $0.135$, and $0.074$ for $n=1/2$, $1$, $3/2$, and $2$, respectively. Because the compactness plays a key role in Eqs.~\eqref{eq:alg_i}, \eqref{eq:alg_l}, and \eqref{eq:alg_q}, differences in the surface values of these functions are not the key factor for those in the second-order moments, $\overline{I}$, $\Lambda$, and $\overline{Q}$.

As the configuration becomes relativistic with a large value of $C$, the functions $w$, $\tilde{q}$, and $f$ take large values throughout the star. This is consistent with the finding in Sec.~\ref{sec:result_C005} that $w_s$, $\tilde{q}_s$, and $f_s$ are on the order of $\mathcal{O} (C)$. Because we may always multiply these functions by a dimensionless factor such as $r/m$ or $(r-2m)/(m+4\pi P r^3)$, this dependence on $C$ may simply be regarded as the result of our specific parametrization. Still, we believe our parametrization is reasonably straightforward. Meanwhile, $y$ tends to stay closer to its central value, $2$, for the maximum-mass configuration. This reflects the intrinsic difficulty in deforming strongly relativistic, centrally condensed objects.

Comparison of Figs.~\ref{fig:C005} and \ref{fig:Max} reveals that the peaks of these functions shift inward as the compactness increases. Moreover, the profiles of $y$ for $n=1/2$ and (marginally) $1$ even exhibit second extrema near the surface. If the speculation made in Sec.~\ref{sec:result_C005} were the whole story or extended to nonmonotonic behavior in general, the universality should have been lost toward the maximum-mass configuration. This is partly true, as soft equations of state with $n \ge 2$ are found to deviate from the universality observed for stiff equations of state strongly toward the maximum-mass configuration \cite{Yagi:2013awa}. For stiff equations of state, however, the approach to the black-hole limit of $C = 0.5$ reduces the sensitivity of appropriately normalized second-order moments, $\overline{I}$, $\Lambda$, and $\overline{Q}$, to the underlying equation of state \cite{Yagi:2013awa}. To sum up, stiff equations of state appear to realize the universality via the coordination of the flat density profile and the large value of achievable compactness.

Before concluding this study, we comment on the relative importance of $\tilde{q}_s$ and $f_s$ in the high-compactness configuration. For the maximum-mass configuration of $n=1/2$ with $C=0.32$, the first term in the second line of Eq.~\eqref{eq:alg_q} is larger than the second term only by a factor of $\approx 4$. Investigations of other configurations reveal that the second term begins to play a nonnegligible role as the compactness increases. While we may still say that $\tilde{q}_s$ plays a dominant role, our tentative identification of $\tilde{q}$ as the primary variable must be taken with care in the strongly relativistic regime. The relative importance of two terms in the denominator does not vary significantly with respect to the compactness.

\section{Summary} \label{sec:summary}

We present a concise formulation for obtaining second-order moments, $\overline{I}$, $\Lambda$, and $\overline{Q}$, of relativistic stars and investigate their relations. These three moments are obtained by solving six ordinary differential equations, Eqs.~\eqref{eq:ode_p}, \eqref{eq:ode_m}, \eqref{eq:ode_w}, \eqref{eq:ode_f}, \eqref{eq:ode_q}, and \eqref{eq:ode_y}, without explicitly deriving homogeneous solutions. If we change the independent variables from $r$ to $H$, Eq.~\eqref{eq:ode_p} is replaced with Eq.~\eqref{eq:ode_r}, by which the other ordinary differential equations are multiplied. Then, the second-order moments are derived using the algebraic expressions, Eqs.~\eqref{eq:alg_i}, \eqref{eq:alg_l}, and \eqref{eq:alg_q}. The only room for further simplification, if at all possible, would be the unification of Eqs.~\eqref{eq:ode_f} and \eqref{eq:ode_q}.

By identifying the primary variables $w$, $y$, and $\tilde{q}$ determining $\overline{I}$, $\Lambda$, and $\overline{Q}$, respectively, we investigate the universality and deviation of these second-order moments by systematically varying the polytropic indices \cite{Yagi:2013bca,Yagi:2013awa}. We find that the soft equation of state and/or the high compactness make the peaks of these functions inward and the profile nonmonotonic in the outer region. Considering nonuniversal trends typically observed for soft equations of state, we speculate that the nonmonotonic behavior of these primary variables in the outer region introduces the variety to the relations among second-order moments. Stiff equations of state might also have suffered from nonmonotonic profiles in the high-compactness regime, and they appear to be saved by the approach to the black-hole limit aided by a sufficiently large value of achievable compactness. We recall that the universality is very likely to hold within $\sim 1\%$ for realistic relativistic stars such as neutron stars, and our systematic study solely intended exploration of its physical understanding. Extension to nuclear-theory-based or particle-physics-motivated equations of state, including the first-order phase transition, is left for future study.

The simplicity of our formulation for obtaining the spin-induced quadrupole moment may be useful in various directions. Gravitational-wave data analysis may find it computationally convenient when the equation of state is sampled directly \cite{LIGOScientific:2018cki}. Analysis of X-ray emission from rapidly rotating neutron stars might also benefit from our approach, although the oblateness tends to play a more important role \cite{Psaltis:2013zja}. In addition, our formulation could help theoretical researchers who have somehow hesitated to compute the quadrupole moment.

\begin{acknowledgements}
 Koutarou Kyutoku is grateful to Kent Yagi and Hayato Miyazono for valuable and helpful discussions. This work was supported by JSPS KAKENHI Grant-in-Aid for Scientific Research (No.~JP19K14720 and No.~JP22K03617).
\end{acknowledgements}

%

\end{document}